\documentclass[amsmath, amssymb, aps, prl, twocolumn, floatfix, superscriptaddress]{revtex4-1}
\usepackage{natbib}
\usepackage{graphicx}
\usepackage{float}

\newcommand{\SCorr}{C_S}
\newcommand{\HCov}{C_h}
\newcommand{\Hbar}[1]{\overline{h_{#1}}}

\begin{document}
\title{Imaging the stochastic microstructure and dynamic development of correlations in perpendicular artificial spin ice}
% * <nsamarth@psu.edu> 2018-02-06T14:40:30.752Z:
% 
% 
\author{Susan Kempinger}
 \affiliation{Department of Physics, The Pennsylvania State University, University Park, Pennsylvania 16802-6300, USA}
 \author{Robert D. Fraleigh}
 \affiliation{Communications, Information and Navigation Office, Applied Research Laboratory, The Pennsylvania State University, University Park, PA 16802}
 \author{Paul E. Lammert}
 \affiliation{Department of Physics, The Pennsylvania State University, University Park, Pennsylvania 16802-6300, USA}
\author{Sheng Zhang}
 \affiliation{Materials Science Division, Argonne National Laboratory, 9700 S. Cass Avenue, Argonne, IL 60439, USA}
 \affiliation{Advanced Science Research Center, The Graduate Center of the City University of New York, 85 Saint Nicholas Terrace, New York, NY, 10031, USA}
 \author{Vincent H. Crespi}
 \affiliation{Department of Physics, The Pennsylvania State University, University Park, Pennsylvania 16802-6300, USA}
 \author{Peter Schiffer}
 \affiliation{Department of Applied Physics and Department of Physics, Yale University, New Haven, CT 06520  USA}
 \author{Nitin Samarth}
 \email[]{nsamarth@psu.edu}
 \affiliation{Department of Physics, The Pennsylvania State University, University Park, Pennsylvania 16802-6300, USA}
 \date{\today}
 \begin{abstract}

We use spatially resolved magneto-optical Kerr microscopy to track the complete microstates of arrays of perpendicular anisotropy nanomagnets during magnetization hysteresis cycles. These measurements allow us to disentangle the intertwined effects of nearest neighbor interaction, disorder, and stochasticity on magnetization switching. We find that the nearest neighbor correlations depend on both interaction strength and disorder. We also find that although the global characteristics of the hysteretic switching are repeatable, the exact microstate sampled is stochastic with the behavior of individual islands varying between nonminally identical runs.  

\end{abstract}   
 \maketitle

Artificially structured lattices have become increasingly popular platforms for studying complex collective phenomena in condensed matter. Examples include artificial graphene\cite{N??dvorn??k2012}, artificial skyrmion lattices\citep{Gilbert2015}, and artificial spin ices\citep{Wang2006, Zhang2012a, Heyderman2013a,Gilbert2016,Morgan2012,Wysin2013,Gilbert2015a,Li2018}. Such artificial lattices are useful because they allow systematic engineering and tuning of properties such as interaction strengths and various types of defects to a degree far exceeding what is possible with naturally occurring crystalline lattices. Studies of artificial spin ice in particular have led to the observation of magnetic monopoles and Dirac strings in well-controlled frustrated geometries \citep{Ladak2011, Mengotti2011,Phatak2011,Mol2009,Perrin2016}, and they have also allowed access to the effects of thermal fluctuations\citep{Kapaklis2014,Farhan2013} and disorder\cite{Budrikis2014, Kohli2011a}. Perpendicular artificial spin ice systems\citep{Mengotti2009, Zhang2012a} are particularly propitious in this context because polar magneto-optical Kerr effect (MOKE) microscopy allows complete \textit{in situ} imaging of microstates and their evolution as an applied field is varied\citep{Fraleigh2017}. 
We have used MOKE microscopy to obtain a microscopically detailed picture of the hysteretic magnetization reversal process and the development of correlations during that process, in both frustrated and unfrustrated arrays. 

Most studies of the hysteresis loops of artificial spin ice systems focus on the {\it macrohistory} of an array, the development of the macrostate, characterized by aggregate quantities, which is reproducible from one field cycle to another. 
This includes the hysteresis curves themselves, equivalent to the raw distribution of switching fields, as well as local switching field distributions accounting for the magnetic fields 
produced by nearby islands, and the development of nearest-neighbor spin correlation as the average
magnetization of the array varies through a field sweep.
Interaction between islands makes a significant and identifiable contribution to the width of 
the raw switching field distribution. %, and creates nearest neighbor correlations up to 0.15 which 
%exhibit hysteresis in the sense of peaking somewhat after the point of zero magnetization.
In this paper, we also focus on the {\it microhistory} of an array, the evolution of its microstate during a field sweep, which is not reproducible from sweep to sweep. Although the energy scale of ambient temperature is very small compared to relevant magnetic energies in these systems, the origin of this stochasticity may be associated with thermal fluctuations that become significant near the coercive field.%; however, our quantitative characterization provides valuable clues. 
%The switching field of an average island varies from run to run with a standard deviation ranging from about 10 Gauss to about 20 Gauss as interaction strength increases, and these 
%variations are significantly correlated for nearest-neighbor islands, with a covariance 
%up to 30 G$^2$. 
%Overlap between the zero-magnetization microstates of an array in different field sweeps 
%gives another window on the microhistory stochasticity. It ranges from 85 to 90 percent
%with no clear dependence on interaction strength.

%%%%%%%%%%%%%%%%%
%%%%% END INTRO

The samples studied in this paper were patterned using electron beam lithography, with a standard liftoff of bilayer poly(methyl methacrylate)/polymethylglutarimide (PMMA/PMGI) resist stack. All samples considered contain frustrated (kagome and triangular) and non-frustrated (hexagonal and square) arrays, with lattice spacing ranges of $600 - 1000$ nm (sample 1) or $500 - 800$ nm (samples 2 and 3). The islands are $400 - 450$ nm in diameter, as confirmed by scanning electron microscopy. Magnetic films of Ti(2 nm)/Pt(10 nm)/[Co(0.3 nm)/Pt(1 nm)]$_8$ were deposited using DC sputtering at Argonne National Lab. We used superconducting quantum interference device (SQUID) magnetometry to confirm the strong perpendicular magnetic anisotropy of these films, as well as to measure the saturation magnetization for each film. Specific details on island size and magnetization properties for the samples considered are found in Table \ref{ProprtiesTable}.

\begin{figure}[h]
\includegraphics[width=\columnwidth]{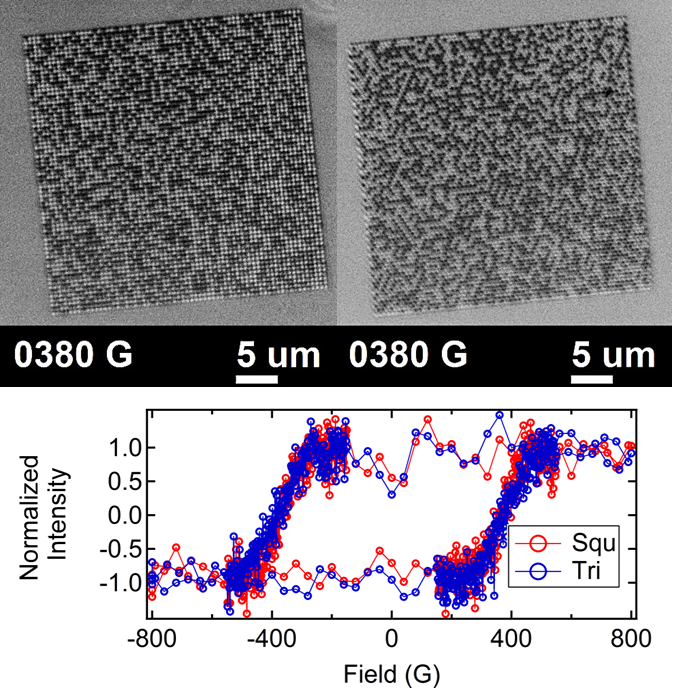}
\caption{Top panel shows MOKE images recorded at 380 G in an increasing the field sweep, near the coercive field, for 500 nm lattice spacing square (left) and triangular (right) arrays from sample 2. The bottom panel shows normalized hysteresis loops recorded using imaging MOKE for these arrays with intensity averaged over the entire array area.}
\label{MOKE}
\end{figure}

Data are collected using an optimized polar MOKE imaging set up, described in detail 
elsewhere\citep{Fraleigh2017}. 
Using image processing techniques, we can resolve, {\it in situ}, 
the magnetization states of every island in an array as shown in Fig.~\ref{MOKE},
thereby obtaining the complete microhistory of the array during a field sweep.
Since each island reverses magnetization only once during a field sweep, a microhistory 
$\alpha$ is encapsulated by the list of switching fields of the islands; 
the value of $H_{\text{app}}$ at which island $i$ switches in sweep $\alpha$ is denoted $h_{i}^{\alpha}$.
Although we do not distinguish notationally, it is to be understood that up-sweeps and down-sweeps
are treated separately, not combined in aggregate quantities or directly compared via correlation
functions.

%%%%%%%%%%%%%%%%%%
\begin{table*}[t]
\begin{center}
\begin{tabular}{c|c|c|c|c|c|c}
\hline\hline
& Diameter (nm) & $M_s$ (A$\slash$m) & $B_0$(500 nm) (G) & $\sigma_d$ (G)& $\sigma_h$ (G) & Avg. Overlap (\%) \\
\hline 
Sample 1 & 400 & 3.46$\times10^5$ & 3.61 & 15.70 & --& --  \\
\hline
Sample 2 & 450 & 3.75$\times10^5$ & 4.96 & 28.21 & 10.8 $\pm$ 1.8 & 87.7 $\pm $1.1  \\
%& .115 $\pm$ .009
\hline 
Sample 3 & 425 & 3.46$\times10^5$ & 4.09 & 17.28 & 9.8 $\pm$ 0.9 & 84.3 $\pm$ 0.8  \\
%& .144 $\pm$ .006
\hline\hline
\end{tabular}

\end{center} 
\caption{Physical, magnetic, and statistical properties of three different artificial spin ice samples}
\label{ProprtiesTable}
\end{table*}
%%%%%%%%%%%%%%%%%

%%%%%%%%%%%%%%%%%%%%%
We begin with the run-to-run consistent, macroscopic (aggregate) aspects, starting with the switching field distribution
and the contribution of island interactions thereto.
The total field experienced by an island comprises not just the externally applied field 
$H_{\text{app}}$, but also a configuration-dependent contribution from other islands which
broadens the distribution of observed (raw) switching fields. 
Without knowledge of the microstates the semi-empirical equation\citep{Fraleigh2017} 
\begin{equation}
	\sigma = A K B_0(L)+\sigma_d
\label{Paper1eqn}
\end{equation}
allows the observed width of the switching field distribution 
to be separated into contributions of island interactions and static disorder, 
the latter presumably introduced by the lithography process.
Here, $A$ is a constant, $K$ an effective coordination number, 
$B_0(L)$ the dipolar field of an island on its nearest neighbor at lattice spacing $L$, 
and $\sigma_d$ the static disorder. 
Additionally, provided the microhistory, we can directly calculate the 
$r$-neighborhood-corrected switching field 
$h_{i,r} = H_{\text{app}} + $ (field from up-to-$r$\textsuperscript{th} neighbors)
in point-dipole approximation, accounting for both the internal and 
external fields felt by an island when it switches. 
In this enriched notation, the raw switching field for sweep $\alpha$ is denoted as  $h_{i,0}^\alpha$.
%%%%%%%%%%%%%%%%%
%%%PEL: I think the following can be left out.
% Using the approximation that islands behave as point dipoles, the field of any island on another is $B(r)=\frac{-\mu_0}{4\pi}\frac{m}{|r|^3}$, which can be calculated from the known geometry and measured magnetization density of the island material. 

%%%%%%%%%%%%%%%%%%%

\begin{figure}[h]
\includegraphics[width=\columnwidth]{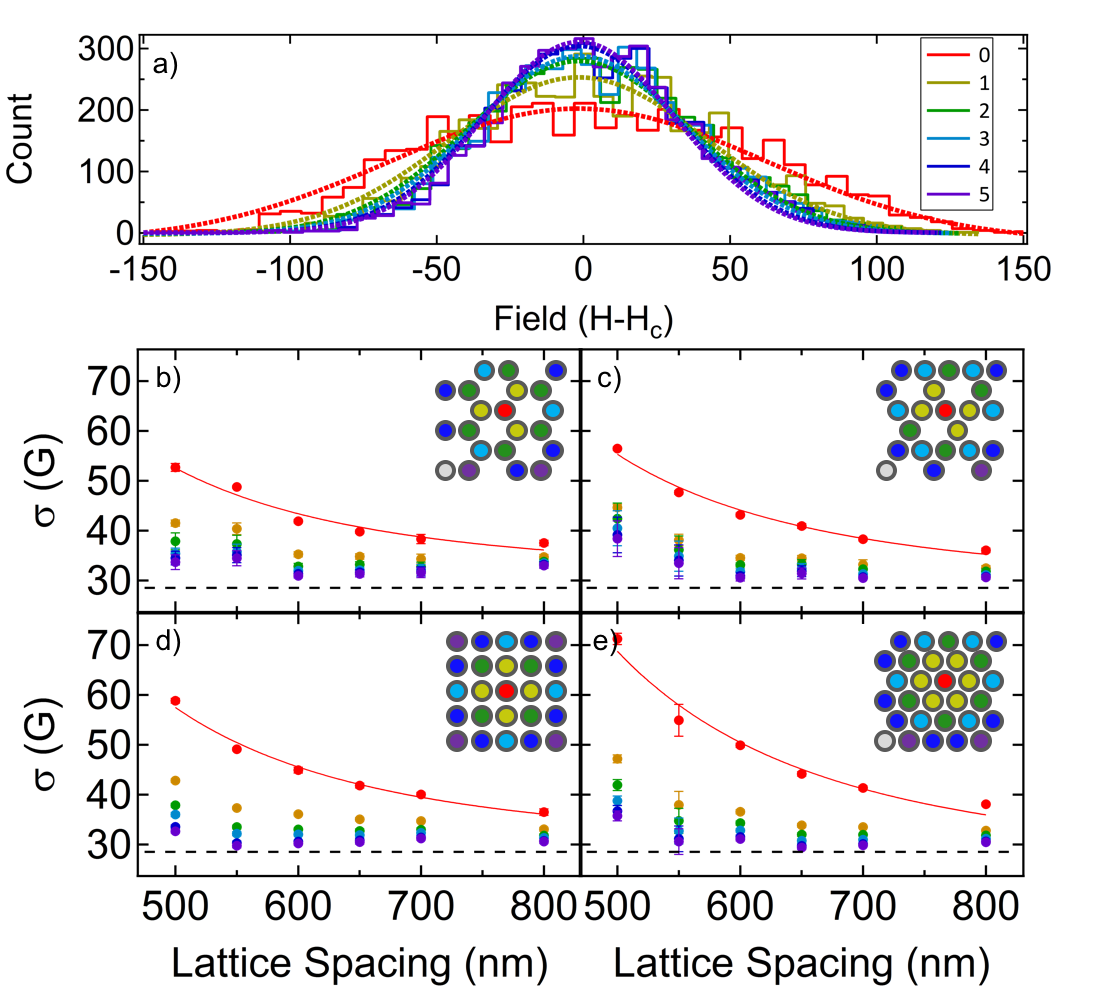}
\caption{a) Switching field distribution and associated Gaussian fits, with switching fields calculated by removing dipolar effects from 0 (as measured) to 5 nearest neighbors for a 500 nm square array (Sample 2). Width of the Gaussian fits for b) hexagonal, c) kagome, d) square, and e) triangular  as a function of lattice spacing taking into account increasing numbers of neighbors. Fits to Eq \ref{Paper1eqn} are shown as red lines, and disorder values from these fits are shown as black dashed lines. The inset images show a cartoon of the lattice geometry colored by target island (red) and neighbor number to match the colors on the graphs. A full set of neighbors is shown up to third nearest neighbor, along with a partial set of fourth and fifth nearest neighbors.}
\label{SFDRemoveNeighbors}
\end{figure}

The top panel of Fig.~\ref{SFDRemoveNeighbors} shows the distributions of the
$r$-neighborhood-corrected switching fields $h_{*,r}^\alpha$ for a single sweep
for a 500 nm square array from sample 2 for $0 \le r \le 5$. 
These are the distributions of all aggregated islands, hence
the `$*$' subscript on $h$.
The expected narrowing of the distribution as $r$ increases (further-neighbor fields accounted for)
is prominent.
The lower panels of Fig.~\ref{SFDRemoveNeighbors} show how the widths of the $h_{*,r}^\alpha$ 
distributions change with lattice spacing for different lattice types.
The broadening in the raw ($h_{*,0}^\alpha$) distributions for different geometries is accounted for completely by the difference in effective coordination number.
For each geometry, as r is increased, the width decreases, becomes independent of the lattice spacing, and approaches the calculated value of disorder. The magnitude of the decrease is on the order of of $A K B_0(L)$, with K calculated considering $r$ neighbors. This decrease agrees with Eq.~(\ref{Paper1eqn}) and reduces to 
the static disorder contribution alone in the limit of large $r$.
This behavior agrees with previous studies pointing to the significance of long-range interactions to the behavior of artificial spin ice\citep{Chioar2014}. The results in Fig.~\ref{SFDRemoveNeighbors} were taken from sample 2; similar results are obtained for sample 3. The analysis supports the treatment of islands as interacting point dipoles, wherein an island's neighbors influence its switching behavior by supplementing the external field with their net dipolar field strength. 

\begin{figure}[h]
\includegraphics[width=\columnwidth]{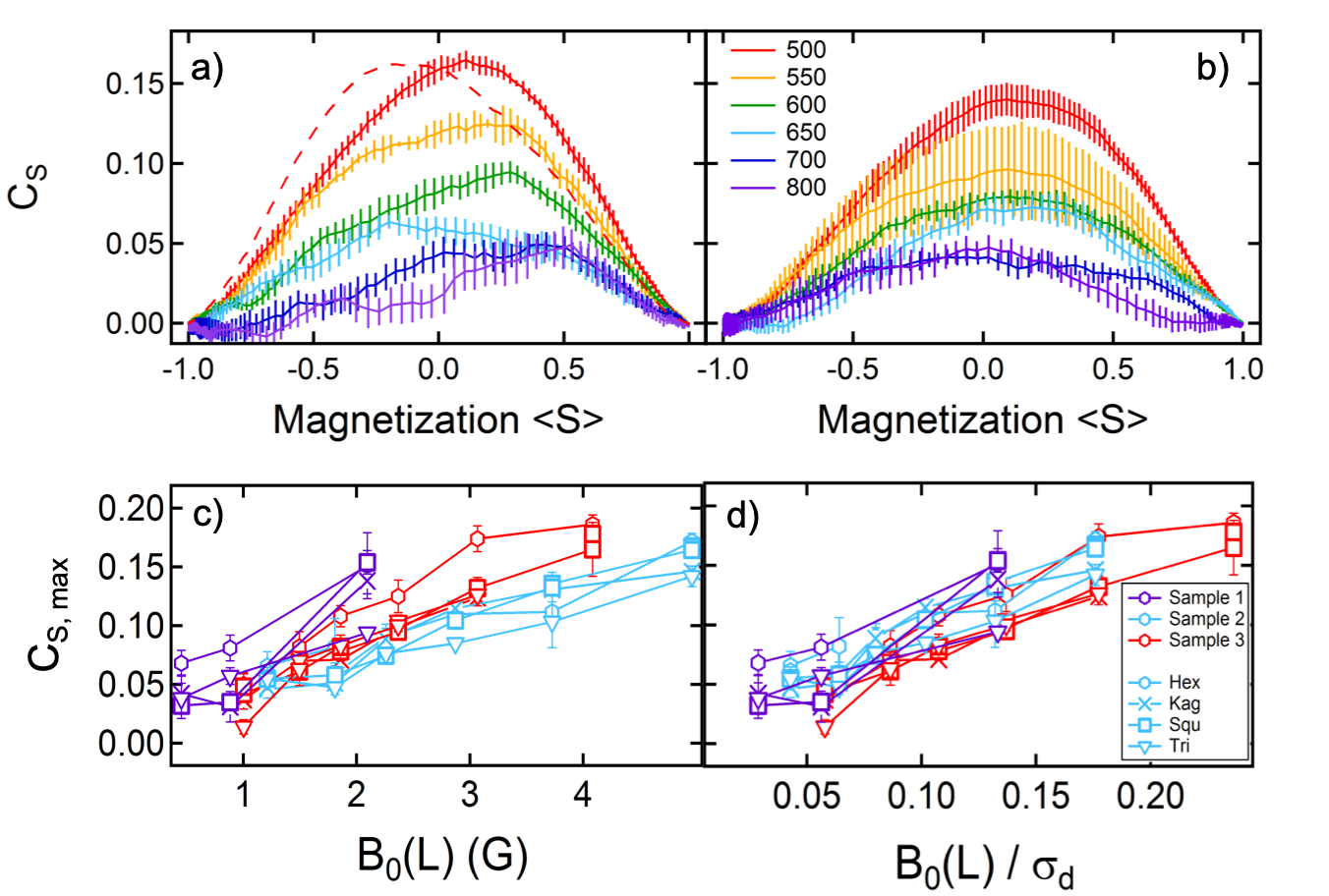}
\caption{Plots of $\SCorr(\langle S \rangle$) for various lattice spacings of a) square, and b) triangular arrays from sample 2, for increasing values of applied field. The dashed line in panel a shows $\SCorr(\langle S \rangle)$ as the field is decreased to more clearly illustrate the asymmetry. Panel c) shows the maximum value of correlation as a function of the dipolar field of an island on its nearest neighbor (i.e. the interaction strength) for samples 1, 2, and 3. Panel d) shows the same data as a function of the dipolar field scaled by the measured disorder in the system.}
\label{Correlations}
\end{figure}

While the $r$-neighborhood-corrected switching field distributions demonstrate an
influence of islands on one another, they say nothing quantitative about correlations. 
We turn to these next.
%%%%%%%%%%%%%%%%%%%
The average spin (magnetization) in an array during sweep 
$\alpha$ is $\langle S_i^\alpha \rangle_i$. Subscripts on averaging brackets 
indicate what is averaged over, and each spin takes value $+1$ or $-1$.
%% SEK these curves are actually from a single run and don't show anything about reproducibility The hysteresis curves in Fig.~\ref{MOKE} show that $\langle S \rangle$ is a fairly
%%reproducible function of external field (for the same sweep direction).
 $\langle S \rangle$ is a fairly reproducible function of external field (for the same sweep direction).
For purposes of comparing different sweeps and different arrays, it is preferable to
parametrize the macrohistory by magnetization $\langle S \rangle$ rather than the applied
field; this will remove fluctuations due to finite size.
Thus, the nearest-neighbor spin correlation for sweep $\alpha$,
% \begin{equation}
\begin{align}
\SCorr^\alpha(\langle S \rangle) &= 
\Big\langle S_i^\alpha \Big\rangle_{i}
\Big\langle S_j^\alpha \Big\rangle_{j}
-\Big\langle S_i^\alpha S_j^\alpha \Big\rangle_{\alpha;\text{NN}}
\nonumber \\
&=
\langle S \rangle^2
-\langle S_i^\alpha S_j^\alpha \rangle_{\text{NN}}
\label{correlation}
\end{align}
% \end{equation}
is regarded as a function of $\langle S \rangle$.
The sum is over all nearest neighbor pairs $(i,j)$ as indicated briefly by the NN subscript. 
$\SCorr$ is zero if spins are independently assigned values $+ 1$ or $-1$ with probabilities 
consistent with $\langle S\rangle$, and we have chosen a sign convention such that it 
increases with the proportion of energetically preferred antiferromagnetic 
nearest-neighbor configurations. 
% Data shown are from the portion of the hysteresis loop with 
% increasing field with exceptions as noted, but the same behavior is observed as the field is decreased.
%%%%%%%%%%%%%%%%%%%
Fig.~\ref{Correlations}a,b show the evolution of $\SCorr^\alpha(\langle S \rangle)$ for 
up-sweeps for the square and triangular arrays on sample 2.
The correlation increases and then decreases as the sample transitions from a saturated state, through zero magnetization to the oppositely saturated state.
However, the correlation does not peak at zero magnetization. Rather, it continues to
increase for a while, peaking at an {\it offset} $\langle S\rangle$. 
This behavior indicates the importance of the quasi-dynamic switching path and the influence 
of island interactions on it.
% , peaking at a value of $\langle S \rangle$ which we call the {\it offset}. 
%In general, we observe that offsets are larger for arrays with smaller interaction strengths where it is more difficult to reach an ordered strength, and arrays with larger interaction strengths which reach more ordered states. The value of static disorder also impacts the offset. For unfrustrated geometries, samples with lower $\sigma_d$ show lower offsets, indicating that the dynamics somehow help the system overcome the disorder. For frustrated samples, there is much less variation in offset between samples, indicating that the dynamics impact frustrated geometries in a qualitatively different way. 
While the offsets are repeatable and observed in multiple samples, the data are too noisy to discern any clear trends in the values. 
%It does seem that the value of static disorder impacts the offset, with samples with lower disorder showing less offset. 

%To further understand how these stepwise effects influence switching, we consider the correlation of field step immediately following $\langle S \rangle $ = 0 on the upward sweep of a hysteresis loop. The exact magnetization of this step depends on how many islands switch, so for simplicity we label it $\langle S \rangle = 0^+$. We find the set of islands with $ h_i=H_c $, and calculate $ \SCorr(\uparrow, 0^+)$. We also consider what would happen if instead of the islands with $h_i =H_c$, a random set of islands switched instead. We restrict the random islands to islands with $\bar{h}_i< H_c + 30$ G to exclude islands whose average field makes is extremely unlikely they would switch near the coercive field. We do this for 1000 sets of islands, and calculate the new values of $ \SCorr(\uparrow, 0^+)$. We subtract this random correlation from the experimental correlation, shown Supporting Information. We observe that for arrays with weak interactions this difference is negligable, but as the interaction strength is increased it becomes statistically distinguishable from zero. This directly demonstrates that stepwise accommodation of local order is enhanced compared to random switching as we increase the interactions.

One anticipates that the maximum value of nearest-neighbor antiferromagnetic correlation 
will increase with the strength of interactions, $B_0(L)$.
Fig~\ref{Correlations}c shows that this expectation is borne out and that the dependence is roughly linear. 
Data for samples 1, 2, and 3 are plotted in different colors.
For each sample, the maximum value of $\SCorr(\langle S \rangle)$ is consistent among all geometries, 
indicating that the interactions are not sufficiently strong for the distinction between frustrated and
unfrustrated geometry to manifest in the macrostate.
However, there is distinct variation in the correlations between samples, indicating that interaction 
strength is an insufficient parameter to characterize these systems. Using instead the dimensionless 
ratio $B_0(L)/\sigma_d$ of interaction strength to static disorder as independent variable, a significant,
albeit partial, data collapse is obtained, as shown in Fig.~\ref{Correlations}d.
Quite reasonably, local ordering is enhanced by increasing interaction strength and hampered by
increasing static disorder.
%% To compare $\SCorr$ across multiple samples, we consider the maximum correlation attained as a function of $B_0(L)$ (proportional to the NN interaction strength). This is shown in Fig \ref{Correlations}c. Each color in the figure corresponds to a different sample. For a given sample, the maximum value of $\SCorr(\langle S \rangle)$ is consistent across all different geometries, which is further evidence that these arrays model behavior for weakly interacting systems and are not yet in a regime where frustration significantly impacts global parameters. However, there is distinct variation in the correlations between samples. This indicates that interaction strength is an insufficient parameter to characterize these systems. Using instead the dimensionless parameter of interaction strength ($B_0(L)$) divided by disorder ($\sigma_d$) as our independent variable, we see a good agreement of behavior between different samples. This parameter indicates that increasing the interaction strength increases the tendency to local ordering, while increasing the disorder decreases the ability of the system to order well.  

Data and analyses discussed to this point show that the systems are macroscopically determinate, in that 
the histories of the global quantities $\langle S \rangle$ and $\SCorr$, as well as distributions of 
switching fields $h_{i,0}$ and $h_{i,r}$, are very similar run-to-run.
%%%PEL: We should have been making this point more explicitly during the course of preceding discussion.
A perfectly deterministic system, though, would have a reproducible {\it microhistory}, 
following exactly the same sequence of island switchings each time it is subjected to 
the same external field sweep. 
Possibly the simplest quantification of nonreproducibility is the run-to-run
switching field variance 
\begin{equation}
\sigma_h^2 % = \var^{\text{r/r}}(h)  
= \Big\langle \Big( h_i^\alpha - \Hbar{i} \Big)^2 \Big\rangle_{i,\alpha},
\label{eq:run-to-run-variance}
\end{equation}
where
\begin{equation}
\Hbar{i} = \langle{h_i^\beta}\rangle_\beta 
\label{eq:run-averaged-h}
\end{equation}
is the run-averaged switching field of island $i$.
The average in Eq.~(\ref{eq:run-to-run-variance}) is over islands in the array and multiple 
(\hbox{7 -- 10}) macroscopically identical hysteresis loops.
In contrast to the {\em aggregate} switching field distributions displayed in 
Fig.~\ref{SFDRemoveNeighbors}, 
the run-to-run variance inherently involves an
average over runs and involves subtraction of an {\em island-dependent} mean. 
Table~\ref{ProprtiesTable} reports average values across all geometries of the run-to-run switching field standard deviation (the square root of the variance) 
$\sigma_h$ for samples 2 and 3 at lattice spacings above 650 nm, of around 10 G. 
The standard deviation increases with increasing interaction strength, maximizing at around 20 G for the arrays with the strongest interaction. 
%It also increases slightly as a function of coordination number, but the change with coordination number is slight.
%These values, of
%around 10 G, are much less than the width of the aggregate switching-field distribution
%because the $\Hbar{i}$ differ.
These values are much less than the width of the aggregate switching field distribution because they are measuring different quantities. The aggregate switching field distribution measures the variation of $h_i$ throughout a lattice, while these values measure the variation of individual island's switching field around it's mean value over a series of distinct runs.
%%% PEL: Should we say something about the distribution of $\Hbar{i}$?

\begin{figure}[h]
\includegraphics[width=\columnwidth]{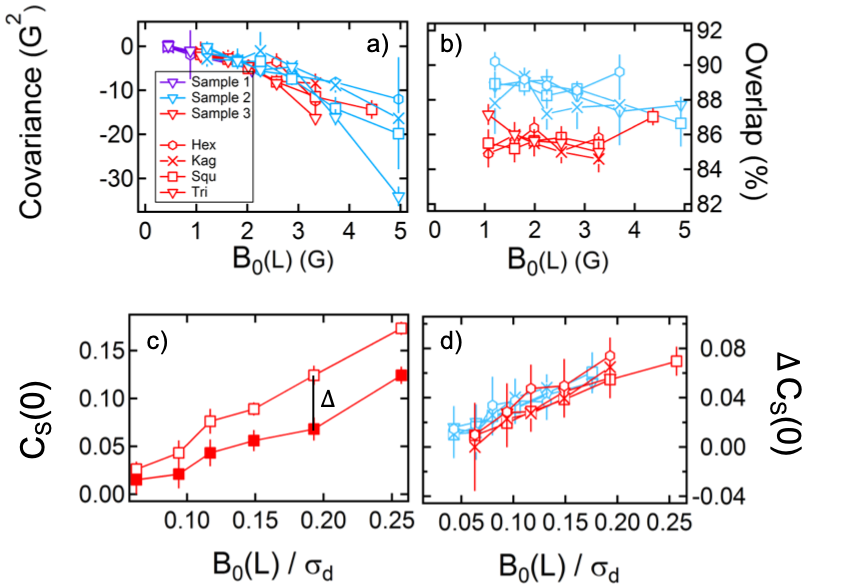}
\caption{a) The covariance of switching fields between multiple runs for all three samples. b) The average overlap at the coercive field between pairs of runs for samples 2 and 3. Overlap is defined as the percent of islands which are in the same state in both states considered. c) The experimentally measured value of $\SCorr(0)$ (open squares) and the average value of $\SCorr(0)$ for randomly generated states with the experimentally measured overlap with the experimental state (solid squares), for a 500 nm square array from sample 3. The difference between these two curves is defined to be $\Delta \SCorr(0)$ d) The average difference in correlation between the experimentally measured state and a state with the experimentally measured average overlap.}
\label{RepeatabilityFig}
\end{figure}

Island switching is significantly influenced by local environment; this is already
clear from the switching field distributions in Fig.~\ref{SFDRemoveNeighbors}.
An indication of how this influence contributes to microhistory variation is provided by
the switching field {\em covariance}
%\begin{align}
\begin{equation}
\HCov  = 
%\mathrm{cov}^{\text{r/r}}(h) =
\Big\langle (h_{i}^\alpha-\Hbar{i}) (h_{j}^\alpha-\Hbar{j}) \Big\rangle_{\alpha;\text{NN}}.
\label{eq:run-to-run-covariance}
\end{equation}
% \nonumber \\ 
% & = 
% \Big\langle 
% (h_{i}^\alpha-\langle h_i^\beta\rangle_\beta) 
% (h_{j}^\alpha-\langle h_j^\beta\rangle_\beta) 
% \Big\rangle_{\alpha;ij\in\mathrm{NN}}
% \nonumber \\ 
% & = 
% \Big\langle 
% \langle h_i^\alpha h_j^\alpha \rangle_{\alpha} 
% - \langle h_{i}^\beta \rangle_\beta \langle h_{j}^\gamma \rangle_\gamma
% \Big\rangle_{ij\in\mathrm{NN}} 
% \end{align}
This quantity is plotted for all arrays in  Fig.~\ref{RepeatabilityFig}a as a function of interaction strength $B_0(L)$.  
That $\HCov$ is negative conforms to expectations since if one island switches ``early'', it will 
increase the energy barrier for a neighbor to switch, due to the antiferromagnetic interactions. The 
arrays with the weakest interaction, although they show significant $\sigma_h$ (Table~\ref{ProprtiesTable}),
show no significant covariance.
%% PEL: "statistically significant" is probably best avoided, as it just invites questions
%% such as "at what level?"
As the interactions are increased, the covariance between neighboring island's switching fields increase in magnitude. 
The increase in covariance also increases as a function of effective coordination number, similar to how the switching field distribution broadens with effective coordination number. In fact, at these interaction strengths, the impact of array geometry can be described completely by the coordination of the array, rather than whether or not there is frustration.
The behavior at low interaction strength gives an indication of the intrinsic behavior of the islands, and the change with increasing interaction strength allows us to judge the impact of interactions.
Because dynamics play a large role in the correlations of these systems, and there is some level of random variation that propagates through the lattice by neighbor interactions, it is likely that we will observe significant differences in the microstates. 

To further characterize (non)reproducibility of the microhistory, we examine the average overlap 
\begin{equation}
\overline{f_{=}}
= \frac{1}{2} \left[ 1 + \left\langle S_i^\alpha S_i^\beta \right\rangle_{i;\alpha\neq\beta} \right] 
\end{equation}
at zero magnetization, $\langle S \rangle = 0$. The average overlap 
is simply the fraction of islands which are in the same state in a randomly chosen pair 
of distinct runs.
Calculated 
values for samples 2 and 3 are plotted in Fig.~\ref{RepeatabilityFig}b and range from
84\% to 90\%.
Sample 2 has a consistently larger overlap than sample 3, which is reasonable since 
$\sigma_h$ is similar for the two samples while $\sigma_d$ is larger for sample 2.
A larger ratio of $\sigma_d/\sigma_h$ implies that each island has access to a smaller 
subset of the switching region, increasing the number of islands in the same state 
at any given point in the switching process. 

%%%%%%%%%%%%%%%
One may wonder whether an average overlap approaching 90\% is enough by itself to 
explain the observed macrohistory repeatability. A simple numerical experiment shows this
is not the case. Starting from one specific $\langle S\rangle = 0$ microstate, we randomly
select a fraction $1-\overline{f_=}$ of islands, flip them, and calculate the change
$\Delta\SCorr(0)$ of the nearest-neighbor correlation (see Fig.~\ref{RepeatabilityFig}c). 
Average values of $\Delta\SCorr(0)$ for 1000 repetitions of this experiment are
plotted in Fig.~\ref{RepeatabilityFig}d. The drop in $\SCorr$ is significantly greater than
the standard deviation of the distribution over runs, hence one concludes that there is more
to the correlations than simply the overlap. Indeed, one may calculate that if
microstate $S'_i$ is obtained from $S_i$ by independently flipping spins with prbability
$1-\overline{f_=}$, that the nearest-neighbor correlation of the new microstate has an expectation value
\begin{equation}
\langle S'_i S'_j\rangle_{\text{NN}} = (1-2\overline{f_{=}})^2 \langle S_i S_j\rangle_{\text{NN}}.
\end{equation}

The origin(s) of microhistory stochasticity are not clear.
Noise arising from the experimental setup, for instance in the power supply or magnet,
seem unlikely to be responsible since such influences would be uniform across the sample;
the magnetic field is quite homogeneous over our small field of view.
%%PEL: can we quantify "quite"?
However, the significant run-to-run switching field covariance shows that 
the stochasticity is at least strongly affected by local conditions. 
{\it Prima facie}, one expects thermal fluctuations to be completely negligible; the energy scale of room temperature $k_B T$ equals the magnetic energy of an island in a field of
order $10^{-1}$ G, about 5\% the field step size, which should lead to a high thermal stability at room temperature. However, near the coercive field, 
thermal fluctuations can be surprisingly significant in understanding the behavior of nanomagnetic systems.\cite{Braun1993, Telepinsky2016}
%one expects thermal fluctuations to be completely negligible; these
%systems are designed to be highly thermally stable at room temperature.
%However, room temperature $k_B T$ equals the magnetic energy of an island in a field of
%order $10^{-1}$ G, about 5\% the field step size, and near the coercive field, 
%thermal fluctuations can be surprisingly significant\cite{Braun1993}.
A non-negligible fraction of islands might be caused to switch in a slightly different field by a thermal fluctuation in a given run, and the ``misstep'' would then be amplified and propagated by island interactions. 
%Possibly the difficult thing to understand then is why the overlap at zero magnetization is so insensitive to interactions.
One might expect these propagated missteps to lead to a decrease in the zero magnetization overlap as the interactions are increased. However, we observe that the overlap is insensitive to interactions. This is possibly because only a subset of islands may be susceptible to thermal fluctuations at any given field step. Any island with a coercivity that is not sufficiently close to a given field is constrained to remain stable in its moment orientation at that field in all runs. 

%%We know the source must be local rather than global as it affects nearby islands differently. This makes instability in the power supply or noise in the magnet unlikely as the magnetic field is quite homogeneous over our small field of view. A more likely explanation is thermal fluctuations. While artificial spin ice materials are designed to be thermally stable at room temperature, this is not necessarily still the case in an external field. The external field reduces the energy barrier for magnetic transitions to a point where, near the transition field for a given island, thermal fluctuations can become significant\cite{Braun1993}. So it is quite possible that, as we sweep the applied field, the variation in switching fields early in the hysteresis process is due to thermal effects. %Then, as we have shown, those variations can propagate through the switching process due to interactions, most likely with additional thermal effects taking place during the hysteresis loop. 

In conclusion,
MOKE microscopy allows us to measure complete microhistories of perpendicular artificial spin ice. 
This level of detail allows a more direct and precise quantification of island interactions
on macrohistory, as for instance in the $r$-neighborhood-corrected switching field distributions
studied here. Perhaps more importantly, it allows precise determination of the degree of
microhistory stochasticity. Our results suggest further studies to examine the mechanisms of microhistory stochasticity and fabrication
of systems with interactions strong enough that frustration effects can be observed. In particular, a study of the impact of thermal fluctuations as a possible source for the stochastic behavior will open the possible inclusion of thermally-induced behavior in a broader range of magnetic systems.
%% In conclusion, by studying hysteresis loops of interacting arrays of magnetic islands, we gain a comprehensive understanding of how interactions affect switching in weakly coupled systems. Using switching field distributions we extract the static disorder strength, and confirm the role of interactions by systematically removing dipolar interactions, leaving only a constant contribution from the static disorder. We then observe how these interactions lead to correlations in the arrays, discerning from the asymmetry in the dynamic correlation curves the importance of dynamics in the development of correlations. While the potential for correlation is determined by the interaction strength and disorder, the point in the switching process at which this correlation is achieved is determined by the dynamics. This allows the system to accommodate different levels of interaction and disorder as well as differences in geometry. There is also a significant stochastic variation from run to run in island switching fields. This leads to large variations in microstate while still maintaining macroscopic repeatability.%, making these systems excellent physical realizations of statistical systems, for which both the complete macroscopic and microscopic properties can be measured. 
%% This variation is surprising, and it's origin remains to be definitively determined. As we did not expect to see dramatic run to run variations, this suggests that there might also be surprising variations in other seemingly stable systems.
 
\begin{acknowledgments}
This project was funded by the US Department of Energy, Office of Basic Energy Sciences, Materials Sciences and Engineering Division under Grant No. DE-SC0010778.
\end{acknowledgments}

%\bibliography{CorrPaper}
%\bibliographystyle{unsrt}
%\bibliographystyle{apsrev4-1.bst}
\bibliographystyle{unsrt}
\bibliographystyle{apsrev4-1.bst}

\end{document}